\documentstyle[psfrag,aps,preprint,epsfig,axodraw]{revtex}

\begin{document}

\tighten

\preprint{TU-653}
\title{Phenomenology of  Minimal Supergravity with Vanishing \\
A and B
Soft Supersymmetry-Breaking Parameters}
\author{Motoi Endo, Motoaki Matsumura and Masahiro Yamaguchi}
\address{Department of Physics, Tohoku University,
Sendai 980-8578, Japan}
\date{July 2002}
\maketitle
\begin{abstract} 
The ansatz of vanishing A and B parameters eliminates CP violating complex 
phases in soft supersymmetry-breaking parameters of the minimal 
supersymmetric standard model, and thus provides a simple solution to the 
supersymmetry CP problem. Phenomenological implications of this ansatz 
are investigated in the framework of minimal supergravity. We show that 
electroweak symmetry breakdown occurs, predicting relatively large $\tan 
\beta$.  The ansatz survives the Higgs mass bound as well as the $b \to 
s \gamma$ constraint if the universal gaugino mass is larger than 300 GeV. 
We also find that the supersymmetric contribution to the anomalous magnetic 
moment of muon lies in an experimentally interesting region of order  
$10^{-9}$ in a large portion of the parameter space.
\end{abstract} 

\clearpage

It is well-known that soft supersymmetry (SUSY) breaking mass
parameters in the minimal supersymmetric standard model (MSSM)
generally have many CP violating phases, which are strongly
constrained by non-observation of the electric dipole moments (EDMs) of
electron \cite{eEDM}, neutron \cite{nEDM} and mercury atom \cite{HgEDM}. 
The hypothesis of the universal gaugino mass and
the universal scalar mass significantly reduces the number of the CP
phases, but there still remain two phases.

In fact, the minimal supergravity model which we will consider here has four
SUSY breaking mass parameters and one SUSY invariant mass:
1) the universal gaugino mass $M_{1/2}$, 2) the universal scalar mass
$m_0$, 3) the SUSY invariant higgsino mass $\mu$, 4) the universal trilinear 
scalar coupling $A$, and 5) the Higgs mixing mass parameter $B\mu$. The two 
physical CP phases can be chosen to be phases of $A M_{1/2}^{*}$ 
and $B M_{1/2}^{*}$, on which stringent constraints are put by the
present upperbounds of the EDMs 
\cite{EDMs1}. (See also {\it e.g.} Ref.~\cite{Abel:2001vy} and 
references therein for a recent analysis.)

One can evade these strong constraints if the $A$ and $B$ parameters
vanish at some high energy scale:
\begin{equation}
  A=B=0.
\label{eq:vanishingAB}
\end{equation}
Renormalization group evolution generates non-zero values for them,
which have the same phase as the gaugino mass.   
In fact, the solution to the SUSY CP problem with the boundary condition
(\ref{eq:vanishingAB}) was considered in gauge mediation of supersymmetry
breaking \cite{Babu:1996jf,Dine:1997xk,Rattazzi:1997fb}.
This solution would be 
also very plausible in gravity mediation, where 
the phase of the gaugino mass is in general  uncorrelated to that of the gravitino mass.
Dynamical  realization of this ansatz in high-scale supersymmetry
breaking scenario has recently been presented 
in~\cite{YamaYoshioka}.

The purpose of this paper is to show that the ansatz (\ref{eq:vanishingAB})
is phenomenologically viable.  To be specific, we take the framework
of the minimal supergravity. We will show that the electroweak
symmetry breaking can take place correctly, yielding relatively large
$\tan \beta$, the ratio of the two Higgs vacuum expectation values
(VEVs). To see viability of the ansatz, we will consider the
constraints from the lightest Higgs boson mass, $b \to s \gamma$, and
the requirement that the lightest superparticle (LSP) be neutral.  It
is found that a wide region of the parameter space survives these
constraints. A special attention is paid to the case where the scalar
mass $m_0$ also vanishes. It turns out that the Higgs mass constraint
will eliminate the parameter region where the LSP is neutral.  We will
summarize several ways proposed to avoid this difficulty.  Finally the
SUSY contribution to the anomalous magnetic moment of the muon will be
briefly
discussed.
 
The model we are considering has, thus, three parameters\footnote{
In the following we take a convention that $M_{1/2}$ and $\mu$ are
real parameters.}  
\begin{equation}
   M_{1/2}, \quad m_0, \quad \mu,
\end{equation}
which are given at some high energy scale. Here we identify it 
with the grand-unification scale of $2 \times 10^{16}$ GeV.  The superparticle
masses at low energy are computed by solving renormalization group equations.
In particular, the values of $A$ and $B$ are given as functions of the 
gaugino mass $M_{1/2}$.  
The condition that electroweak symmetry 
breaking takes place at the correct energy scale gives one relation among
the three parameters. Here we take $M_{1/2}$ and $m_0$ free parameters, and
determine the value of $\mu$. Notice that 
the value of $\tan \beta$ is not an input  parameter, but is rather an output 
when  $m_0$ and $M_{1/2}$ are given. 

\begin{figure}[ht]
  \begin{center}
    \includegraphics[scale=0.45]{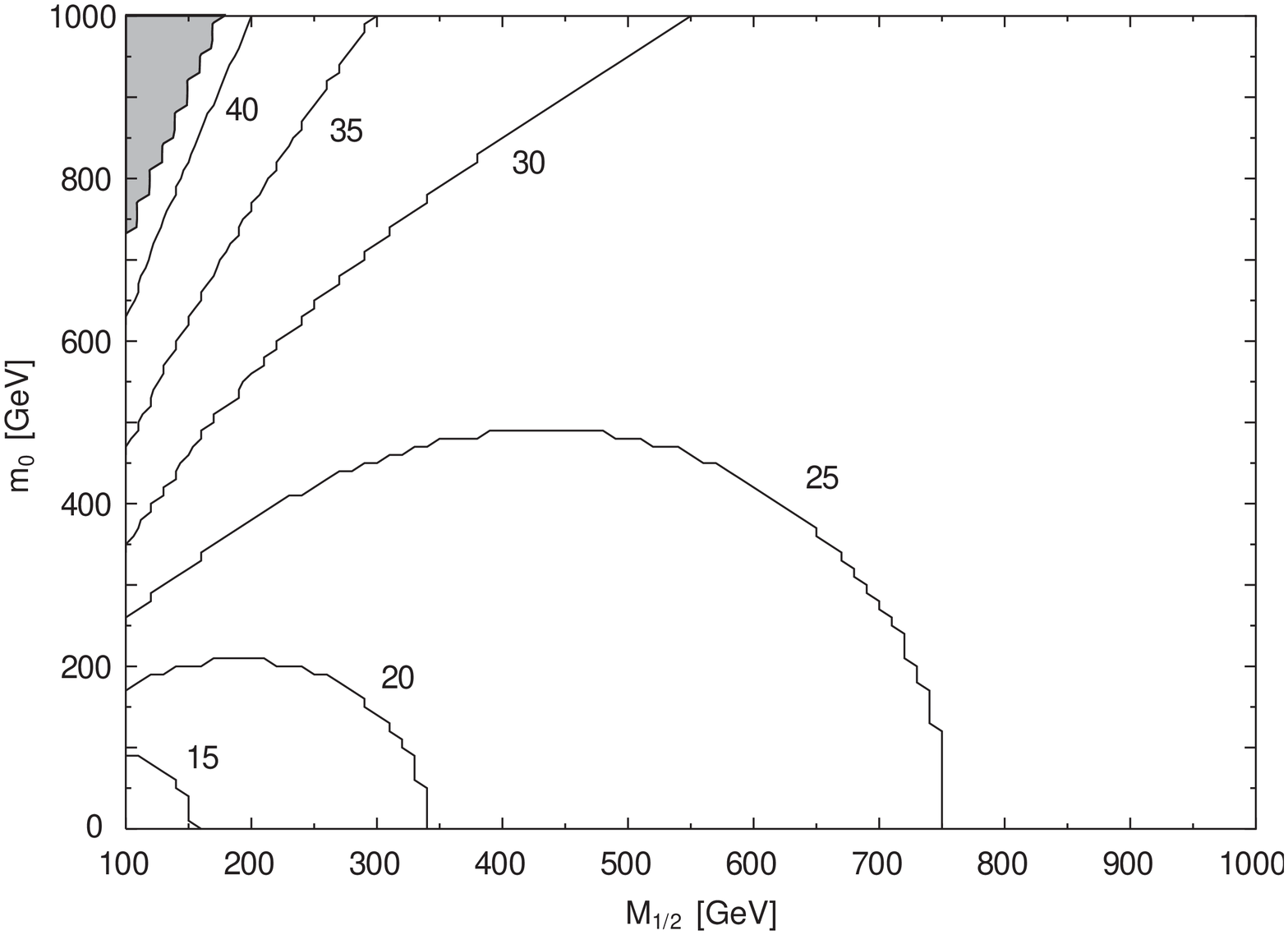}
  \end{center}
  \caption{Constant contours of the value of $\tan \beta$ 
    on $m_0$--$M_{1/2}$ plane in the minimal supergravity model 
    with $A = B = 0$.  The electroweak symmetry breaking does not 
    take place in the shaded region.}
  \label{fig:contour_tanB}
  \begin{center}
    \includegraphics[scale=0.45]{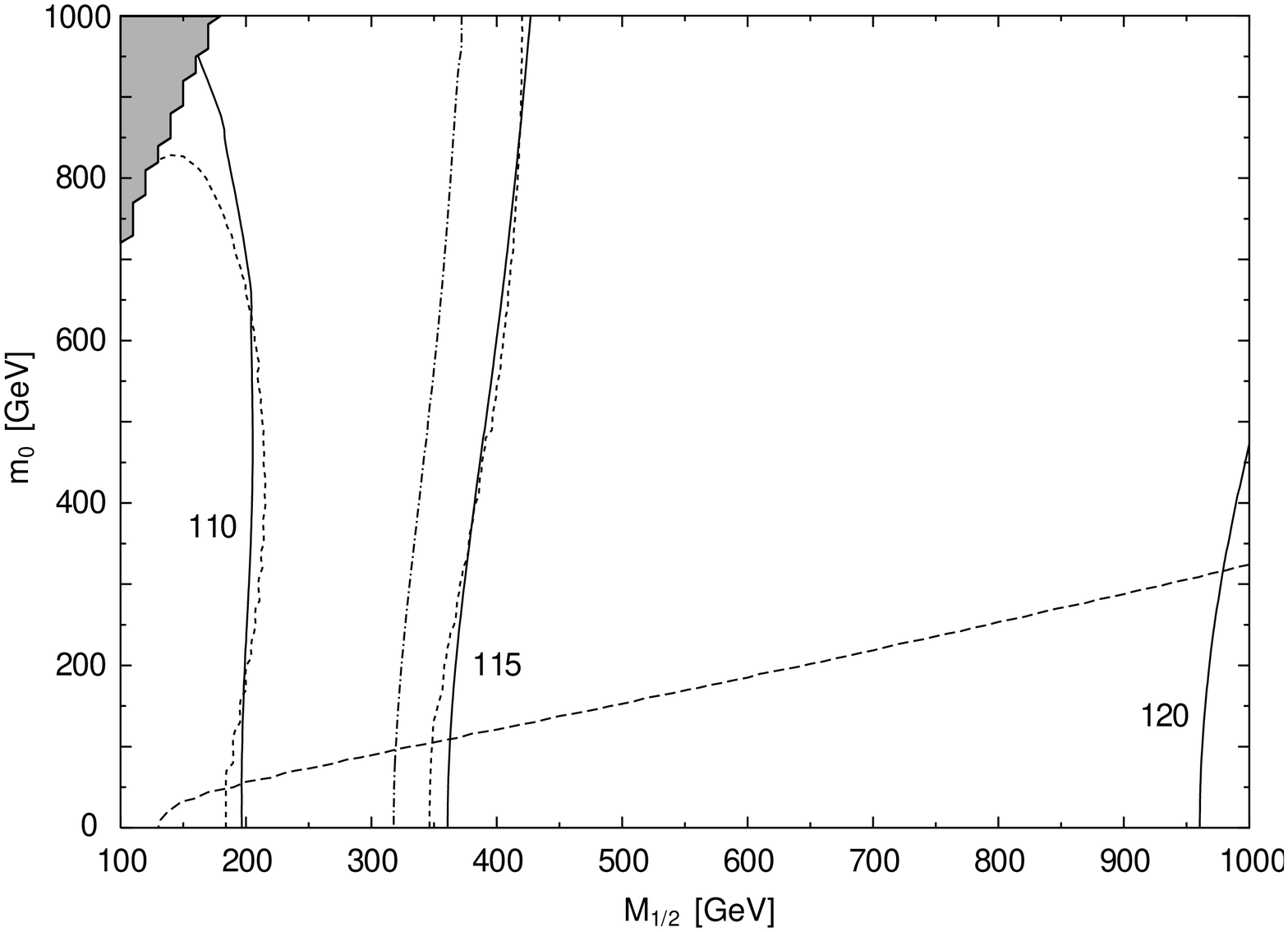}
  \end{center}
  \caption{The experimental bounds on $m_0$--$M_{1/2}$ plane 
    in the minimal supergravity model with $A = B = 0$. 
    The solid lines represent the contours of the constant Higgs boson mass.  
    The dot-dashed line is the present Higgs boson mass bound 
    of $114.1\ {\rm GeV}$.  The dotted lines are the contours of 
    the constant $b \to s \gamma$ branching ratio 
    (${\rm Br}(b \to s \gamma) = 2 \times 10^{-4}$ 
    and $3 \times 10^{-4}$ from left). 
    The stau and the lightest neutralino are degenerate in mass 
    on the dashed line.  The neutralino is the LSP above this line 
    and the stau is lightest below this line. The electroweak symmetry 
    breaking does not take place in the shaded region.}
  \label{fig:contour_bound}
  \begin{center}
    \includegraphics[scale=0.45]{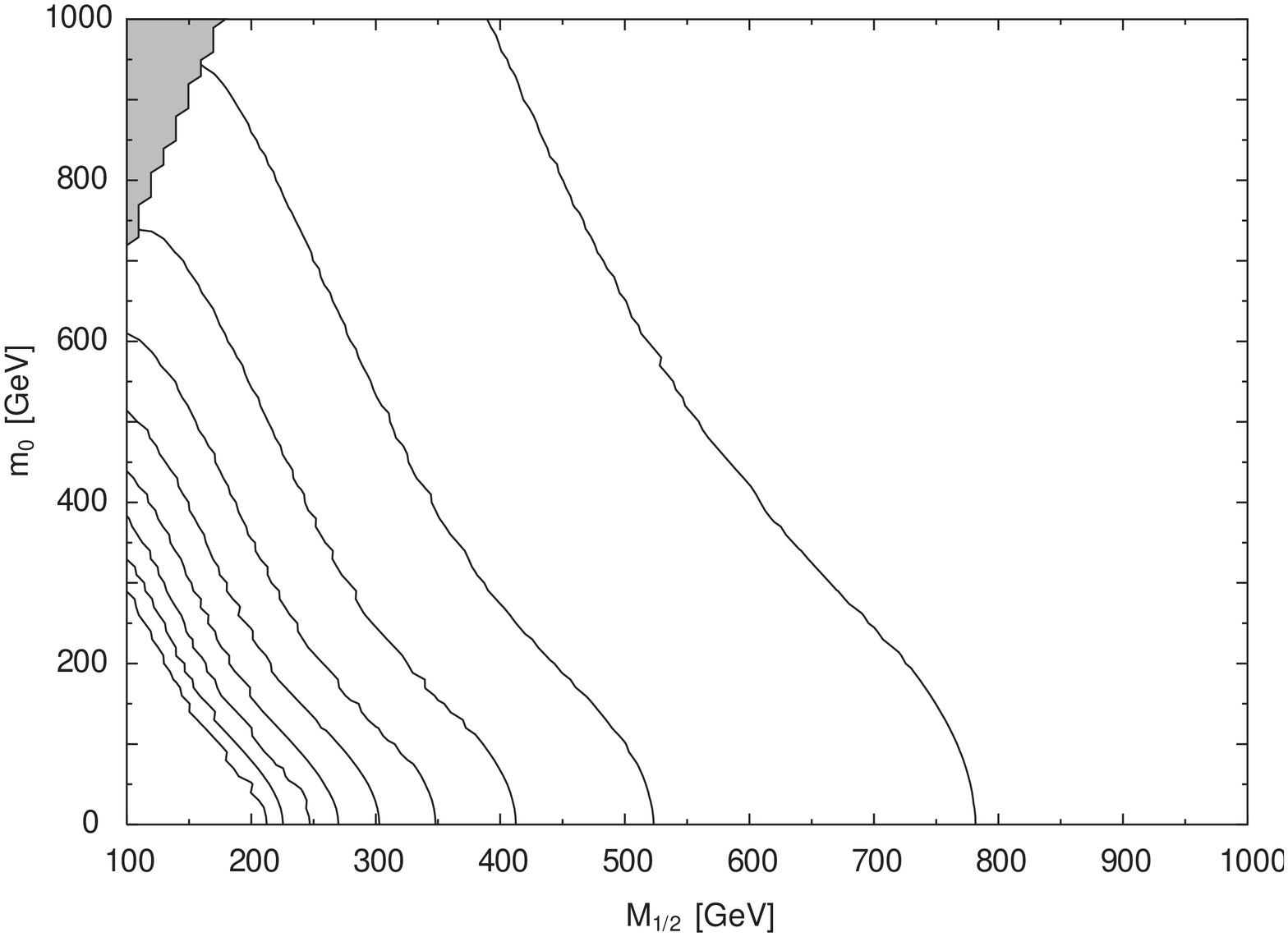}
  \end{center}
  \caption{Constant contours of the value of the SUSY contribution 
    to the anomalous magnetic moment of muon on $m_0$--$M_{1/2}$ plane 
    in the minimal supergravity model with $A = B = 0$.
    The solid lines denote the values of $a_{\mu}(\mbox{SUSY})$ of
    1, 2, $\cdots$, 9 (from right) in units of $10^{-9}$.  
    The electroweak symmetry breaking does not 
    take place in the shaded region.}
  \label{fig:contour_MDM}
\end{figure}

A survey of the parameter space was performed for the range $0 \leq
m_0 \leq 1\ {\rm TeV}$ and $100\ {\rm GeV} \leq M_{1/2} \leq 1\ {\rm TeV}$. 
It turns out that the electroweak symmetry breaking occurs in almost all
region of the parameter space unless $m_{0}$ is much larger than
$M_{1/2}$. 
In the following analysis, we solved the one loop renormalization 
group equations and used the effective potential at one loop order 
to determine the values of $\tan\beta$ and $\mu$.
In Fig. \ref{fig:contour_tanB}, constant contours of the value 
of $\tan \beta$ are plotted in the $m_0$--$M_{1/2}$ plane. 
Here the top quark mass $m_t$ has been fixed to be 174 GeV, {\i.e.}
the central value of the top mass measurements.
 We find that the value of $\tan
\beta$ is relatively large. In particular for $M_{1/2} \gtrsim 300$
GeV, $\tan \beta \sim 20$ -- 35. Furthermore we find a positive
correlation between $\tan \beta$ and $m_0$. It is easily understood
if we recall the relation
\begin{equation}
 \sin 2 \beta  =  -\frac{2 B \mu}{2\mu^2 +\tilde{m}_1^2 +\tilde{m}_2^2},  
\label{eq:sin2beta}
\end{equation}
which is derived by minimizing tree-level scalar potential. Here 
$\tilde{m}_1^2$
and $\tilde{m}_2^2$ are SUSY-breaking mass-squared parameters for the two 
Higgs multiplets $H_1$ and $H_2$. 
As $m_0$ increases, the denominator of (\ref{eq:sin2beta}) increases
and hence $\tan \beta$ also increases.

Here it is interesting to note that the value of the $\mu$ parameter
is determined to be positive in the convention that $\tan \beta$ and
$M_{1/2}$ are taken positive.  In fact, the large top Yukawa coupling
drives the $B$ parameter negative during the renormalization group
flow, which is essential to determine the sign of $\mu$. As we will see
shortly, the sign of this parameter plays an important role when discussing
the constraint from $b \to s \gamma$.

Next we compute the mass of the lightest CP even Higgs boson. 
A contour plot of the Higgs boson mass is given 
in Fig. \ref{fig:contour_bound}.  The present experimental bound 
of 114.1  GeV~\cite{HiggsMass} is also indicated in the same figure. 
To compute the Higgs mass,  we used the {\tt FeynHiggsFast} 
\cite{FeynHiggsFast}.  One finds that the value of the Higgs mass is 
sensitive to the gaugino mass, and the region $M_{1/2} \gtrsim$ 300 GeV 
survives the present Higgs mass bound. 

We also consider the constraint from ${\rm Br}(b \to s \gamma)$.
Since $\mu$ is always positive in the case at hand, 
the SUSY contribution to $b \to s \gamma$ partially
cancels the charged Higgs contribution, and thus the deviation from
the Standard Model prediction is small unless the superparticles are
very light. We followed Ref.~\cite{Kagan:1999ym} to estimate the Standard
Model contribution. As for the charged Higgs contribution, we used the 
next-to-leading order calculation~\cite{chargedHiggs}. The superparticle
loops were basically computed at one loop order.  To evaluate these 
contributions, we additionally took into account corrections 
in powers of  $\tan \beta$, which are important for large 
$\tan \beta$ \cite{Degrassi:2000qf}. 
The calculated branching ratio should be compared with the recent 
measurement at CLEO collaboration ${\rm Br}(b \to s \gamma)=(3.21 \pm 0.43 \pm 
0.27^{+0.18}_{-0.10}) \times 10^{-4}$, where the errors are of statistical,
systematic and theoretical, respectively~\cite{Chen:2001fj}.
Here we take rather a conservative range
\begin{equation}
       2 \times 10^{-4} < {\rm Br}(b \to s \gamma) <4.5 \times 10^{-4}.
\end{equation}
The allowed region consistent with the experimental data 
is  shown in Fig. \ref{fig:contour_bound}. 
One finds that it does not severely restrict the parameter space. In fact, 
once the Higgs mass constraint is imposed,
the $b \to s \gamma$ does not further constrain the parameter space. Namely
the severest constraint comes from the Higgs boson mass. This is 
an interesting characteristic of the $A=B=0$ ansatz. 

Let us next discuss which particle will be the LSP. As is well-known,
cosmological argument\footnote{See for example Ref. \cite{Kudo:2001ie} 
for a recent argument.}
requires that the LSP must be neutral as far as
R-parity is conserved and thus the LSP is stable. In our case, the
lightest neutralino is always bino-like and the LSP (among the MSSM
superparticles) is either the neutralino or a stau. 
In Fig. \ref{fig:contour_bound}, we show
the line where the stau and the lightest neutralino are degenerate in
mass. The region above this line, the neutralino will be the LSP and
thus the region is cosmologically viable. On the other hand, in the
region below this line, the stau becomes the lightest among the
superparticles in the MSSM.

A special attention should be paid to the case of $m_0=0$, which is
often referred to as the no-scale boundary condition. In fact, if
supersymmetry is broken in the sequestered sector where the K\"{a}hler
potential is of the sequestered form, one obtains $m_0=0$ as well as
$A=0$ \cite{Randall:1999uk,Luty:2000cz} (see also Ref. \cite{Inoue:1992rk}).  
The boundary condition $m_0=A=B=0$ was considered in the framework of
gaugino mediation~\cite{Schmaltz:2000gy}.
When $m_0=0$, the stau mass
and the bino-like neutralino mass are very degenerate. It is known
that the neutralino is lighter only when $M_{1/2}\lesssim 300$ GeV,
which corresponds to $M_1 \lesssim $ 120 GeV\cite{Inoue:1992rk}.  
The situation becomes
even worse when $\tan \beta$ is large, which is indeed the case in our
ansatz, because the stau becomes light partly due to the
renormalization group effect coming from the non-negligible tau Yukawa
coupling and partly due to the left-right mixing of the stau
mass-squared matrix.  From Fig. \ref{fig:contour_bound}, 
one sees that the neutralino is lighter than the stau 
only if $M_{1/2} \lesssim 130$ GeV. As was discussed earlier, 
this region is already excluded by the Higgs mass
bound which requires $M_{1/2} \gtrsim 300$ GeV.

One might think that one can exclude the no-scale boundary condition
$m_0=0$ in our case. However this is not necessarily true. There are
several ways out proposed in the literature. Firstly if the theory is
embedded in a grand unified theory (GUT), the renormalization group
effect above the GUT scale can sufficiently raise the stau mass,
because the right-handed stau is in the 10-plet of 
SU(5) \cite{Kawamura:1995ys,Schmaltz:2000gy}.\footnote{
The renormalization group flow above the GUT does not significantly change the
other conclusions about the Higgs mass bound and the $b \to s \gamma$ 
constraint.}
  Secondly
D-term contribution to the stau can give a positive correction to the
stau mass \cite{Fujii:2002mb}. Thirdly the gravitino can be 
lighter than the stau. In this
case the stau decays to gravitino and thus is unstable.
\footnote{A cosmological
constraint may come from the requirement that the stau decay does not 
spoil the success of big-bang nucleosynthesis 
\cite{Moroi:1993mb,Gherghetta:1999tq,Asaka:2000zh}.}

We also would like to briefly discuss SUSY contribution
to the anomalous magnetic moment of muon $a_{\mu}(\rm{SUSY})$. 
Since $\tan \beta$ is
large, the SUSY contribution is quite sizable. In Fig. \ref{fig:contour_MDM}, 
we draw constant contours of the values of
$a_{\mu}(\rm{SUSY})$. In a large portion of the parameter space, it is 
of the order of $10^{-9}$, which may be accessible in near future
experiments. 

Finally we should note here how our results suffer from the uncertainty of  
the top quark mass. 
In fact, larger top Yukawa coupling makes the magnitude of $B$ 
parameter larger, and also enhances the radiative correction 
to the lightest Higgs boson mass.  
Therefore, $\tan\beta$ becomes smaller and the constraint from 
the Higgs boson mass becomes looser when the top quark mass increases. 
We analyzed the case where  $m_t = 179\ {\rm GeV}$ 
which is  1-$\sigma$ away from the central experimetal value of the top 
quark mass.  
In this case, $\tan\beta$ becomes smaller than about $2-3$ compared to 
the previous case of $m_t = 174\ {\rm GeV}$.  
And the Higgs boson mass bocomes larger about $1-3\ {\rm GeV}$.
Thus we conclude that our results are rather insensitive to the change of
the top quark mass.

Before closing, we should emphasize that the ansatz presented here has
a characteristic feature for the superparticle masses and can be
tested in future collider experiments. In this respect, it may be
very interesting if one will be able to reconstruct the Higgs potential
by using experimental data, which may reveal how the SUSY CP problem is
solved in nature.

\section*{Acknowledgments}
We would like to thank M. Kakizaki and K. Yoshioka for useful discussions.
This work was supported in part by the Grant-in-aid from the Ministry
of Education, Culture, Sports, Science and Technology, Japan
(No. 12047201).

\end{document}